\documentstyle[aas2pp4]{article}

\begin{document}
\title{First Results from the All-Sky Monitor on the Rossi X-ray Timing Explorer}\author{Alan M. Levine\altaffilmark{1}, Hale Bradt\altaffilmark{1}$^,$\altaffilmark{2}, Wei Cui\altaffilmark{1}, J. G. Jernigan\altaffilmark{3}, Edward H. Morgan\altaffilmark{1}, Ronald Remillard\altaffilmark{1}, Robert E. Shirey\altaffilmark{1}$^,$\altaffilmark{2}, Donald A. Smith\altaffilmark{1}$^,$\altaffilmark{2}}
\altaffiltext{1}{Center for Space Research, Massachusetts Institute of Technology, Cambridge, MA 02139}
\altaffiltext{2}{Department of Physics, Massachusetts Institute of Technology, Cambridge, MA 02139}
\altaffiltext{3}{Space Sciences Laboratory, University of California, Berkeley, CA 94720}
\authoremail{aml@space.mit.edu}

\begin{abstract}
The All-Sky Monitor on the {\it Rossi X-ray Timing Explorer} has been
monitoring the sky in the 1.5 - 12 keV band since late February.  The
instrument consists of three coded-aperture cameras which can be
rotated to view different regions by a motorized drive assembly.
Intensities of $\sim$100 known sources are obtained via least-square fits 
of shadow patterns to the data and compiled to form x-ray light curves.  
Six orbital periodicities and four long-term periodicities, all previously 
known, have been detected in these light curves.  
Searches for additional sources have also been conducted.  X-ray
light curves for the Crab Nebula, Cyg X-1, 4U 1705-44, GRO J1655-40,
and SMC X-1 are reported. They illustrate the quality of the results
and the range of observed phenomena.  
\end{abstract}


\section{Introduction}

The All-Sky Monitor (ASM) on the {\it Rossi X-ray Timing Explorer} ({\it RXTE})
is now monitoring the highly variable and often unpredictable X-ray sky.
In the tradition established by the all-sky monitors on the {\it Vela 5B},
{\it Ariel V}, and {\it Ginga} satellites \markcite{holt1987} (Holt \& 
Priedhorsky 1987, \markcite{pried1987} Priedhorsky \& Holt 1987, 
\markcite{tsunemi1989} Tsunemi et al. 1989), and more recently by the BATSE
instrument on the {\it Compton Gamma Ray Observatory} \markcite{fish1989} 
(Fishman et al. 1989), our results are being used to alert observers to the 
appearance of transients and to other time variable phenomena and to record
long-term intensity histories of bright X-ray sources. Herein we
briefly describe the instrument and data analysis and illustrate the
results with selected light curves.

\section{Instrumentation}

The ASM consists of three Scanning Shadow Cameras (SSCs) mounted on a
motorized rotation drive. Each SSC contains a position-sensitive
proportional counter (PSPC) that views the sky through a slit mask
(Fig. 1). The PSPC is used to measure the displacements and
strengths of the shadow patterns cast by X-ray sources within the
field of view (FOV), and to thereby infer the directions and
intensities of the sources.

The mask is a thin aluminum sheet which is subdivided into $6 \times 2$
subsections. Each of the 12 subsections contains $\sim$15 open and $\sim$16
closed slit elements of size 1 mm by 110 mm arranged in one of 6
carefully chosen pseudo-random patterns.  The 1-mm width of an element
near the center of the mask subtends 12$\arcmin$ at the PSPC.

The volume between the slit mask and the PSPC is divided into two
halves by a partition so that each half of the PSPC views the sky only
through half ($6 \times 1$ subsections) of the mask.  This yields a
FOV of 6$\arcdeg$ by 90$\arcdeg$ FWHM ($12\arcdeg \times 110\arcdeg$ FWZI).

Each PSPC contains 8 resistive carbon-coated quartz fiber anodes, each
end of which is connected to a dedicated electronic measurement chain.
An X-ray event detected in the volume surrounding one of these anodes
yields a pair of pulse heights, which are used to compute the event
energy and position (i.e., the coordinate perpendicular to the slits
in the mask) via the charge-division technique. The achieved position
resolution is 0.2 to 0.5 mm, depending on photon energy. Each PSPC
also contains 12 metal anodes which are used to veto events caused by
charged particles.

X-rays must penetrate through an $8 {\mu}\rm m$ thick aluminized
plastic thermal shield and a $50 {\mu}\rm m$ thick beryllium window to
enter the PSPC.  The PSPCs in SSCs 2 \& 3 also have a thin ($\sim 2
{\mu}\rm m$) coating of polyimide on the interior of the beryllium
foil for protection against possible leaks.  The X-rays are absorbed
in a 95\% xenon - 5\% CO$_2$ gas mixture (total pressure 1.2
atm). Each SSC is sensitive in the energy range of approximately 1.5 -
12 keV, with on-axis effective areas of $\sim$10 cm$^2$, $\sim$30
cm$^2$, and $\sim$23 cm$^2$ at 2, 5, and 10 keV, respectively.

Event data are normally compressed within the two ASM Event Analyzers
(EAs) in the Experiment Data System and relayed to the spacecraft for
insertion in the telemetry stream. One ASM EA accumulates histograms
of counts binned as a function of position for each of the resistive
anodes and for each of three energy bands roughly corresponding to 1.5 -
3, 3 - 5, and 5 - 12 keV.  These position histograms are accumulated in
series of $\sim$90-s ``dwells''. During each dwell, the spacecraft
maintains a fixed attitude and the ASM rotation drive, which is also
controlled by this EA, is not active, so that the orientation of each
SSC is fixed in relation to the sky.  A position histogram thus
contains the superposition of the mask shadows from each X-ray source
in the FOV during a single dwell (Fig. 2).  The start times and
rotation drive angles for each dwell and high voltage on/off commands
are planned in advance on the ground to minimize earth occultations,
and to avoid operation under conditions which may present extra risk
to the PSPCs (see below).

The other ASM EA produces count rates for both X-ray
($\slantfrac{1}{8}$-s time bins) and background (1-s time bins)
events, and pulse height spectra (64-s time resolution, 64 channels
for 0 - 20 keV).

When the high voltage supplies of the SSCs were first activated on
1996 Jan 5, the entire ASM operated normally. However, on the next day
SSC 3 developed high-voltage breakdown to two adjacent carbon-coated
quartz anodes, and on Jan 12, SSC 2 developed similar breakdown on one
anode. All three PSPCs were then turned off so the problems could be
investigated. No cause for the breakdown events was identified.
However, procedures were adopted to avoid operations in additional
high background regions, and whenever the sun with its high X-ray flux
could be in the FOV. Both malfunctioning PSPCs were recovered by
letting the breakdown proceed until it largely ceased, probably due to
the discharge eroding the carbon from the entire lengths of the
affected anodes. ``Normal'' operation commenced for SSC 1 with all 8
anodes on Feb. 22, for SSC 2 with 7 anodes on Mar. 12, and for SSC 3
with 6 anodes on Mar. 19. Since these dates, the operation has been
essentially continuous except for a few days when operational
difficulties intervened. Currently, the ASM collects useful data with
a duty cycle of $\sim$40\%. This duty cycle, together with spacecraft
manuevers which are planned to carry out the observing program of the
other {\it RXTE} instruments, produces a highly stochastic pattern of sky
coverage with a randomly chosen source being scanned typically 5 - 10
times per day.

The gain of the three PSPCs has remained stable to $\sim$1\% since
launch, as determined by measurement of 6 keV X-rays from weak
$^5{}^5$Fe calibration sources.

\section{Data Analysis}

The analysis proceeds by computing intensities for sources
listed as active in a master catalog, and then by searching for and
locating additional sources.  The master catalog includes all X-ray
sources with accurate positions (uncertainty $<$ 3$\arcmin$) and recorded
instances of X-ray flux $>$ 3 mCrab at 2-10 keV. Many of these sources
are transient in nature and may be below the ASM detection threshold
for years at a time. Therefore, the catalog flags as active only the
sources we may expect to detect with the ASM. This helps avoid
needless problems with source confusion and computation time.

Source intensities are obtained from the solution of a linear
least-squares fit of position histograms with model shadow patterns
for each active source within the field of view and with patterns
representing non-X-ray and diffuse X-ray backgrounds (cf. Doty 1988).
The amplitudes of the patterns are taken as free parameters. First, an
unweighted fit is computed. The fit is then iterated twice using as
weights the reciprocals of the variances of the counts per bin
expected from the solution computed in the previous iteration. The fit
solution also yields estimates of the uncertainties of the derived
intensities; these uncertainties are based purely on the photon
counting statistics predicted by the best-fit model.

Two types of problems with the analysis are currently addressed as
follows. First, the fit is repeated with sources whose fitted
intensities are below two standard deviations being removed one at a
time until no more such intensities are obtained from the fit. This is
done to eliminate negative fitted source intensities from the model,
and also to help reduce problems of confusion in crowded fields. The
fitted intensity and the associated statistical uncertainty for each
source is saved from the last solution in which it is listed.  Second,
the fitting process is repeated with small adjustments being made to
the pointing direction of the SSC, since calibration of the SSC
pointing directions and rotation drive angle sensors is not yet
sufficiently accurate for the most demanding cases (e.g., with Sco X-1
in the field). Source intensities and errors (and estimated
adjustments to the pointing direction) are reported from the fit with
the minimum value of the reduced chi-square statistic.

The fit residuals (see Fig. 2) are examined via a cross-correlation
technique for evidence of sources beyond those from the active list.
When such a source is detected, the cross-correlation function is used
to estimate its celestial location. Since the location derived from a
single dwell of one SSC is long and narrow ($\sim5\arcmin \times
5\arcdeg$ for a $\sim100$ mCrab source), this is best done using the
intersections of error boxes from multiple dwells and SSCs. The
derived location is used to attempt to identify the source with
objects flagged as inactive in the master catalog.  The coordinates
from the analysis or from the catalog listing (if the source was
successfully identified) are entered into the active catalog and the
fit procedure is repeated.

To date, this analysis has been performed on a dwell-by-dwell and
SSC-by-SSC basis, but may be extended to data from multiple dwells and
SSCs.

Fitted intensities are normalized to on-axis count rates in SSC 1.
This requires two corrections, the first of which must be applied
because the position histogram models computed for unit count rate
sources do not yet fully take into account the loss of effective area
for sources at large elevations in the FOV.  At present, this
correction factor has been empirically determined from observations of
the Crab Nebula.  The second correction accounts for absorption in the
polyimide coating on the windows of SSC's 2 and 3, so that the results
from all three SSCs are consistent.

\section{Results}

The fitted intensities depend upon the accuracy of the model of the shadow
patterns and upon spectrum-dependent differences among the SSCs and
individual anodes.  The model patterns are based upon both laboratory
and in-orbit calibrations. There are still small systematic
differences between the measured position histograms and the models
(Fig. 2).  We will continue to refine the models.

The corrected intensities in the 1.5 - 12 keV band from individual
dwells are selected so as to reject results from fits with high values
of reduced chi-square, short exposures due to entry into high
background regions, sources close to or behind the Earth's limb,
sources close to the edge of the FOV, or fields which are
exceptionally crowded.  The remaining intensities are plotted in light
curves of each of $\sim$100 sources.  One-day averages have also been
plotted. Selected light curves are shown in Figure 3.

The intensities for the Crab Nebula have an apparent scatter about the
mean of $\sim$5\% (Fig. 3), of which $\sim$4\% can be attributed to counting
statistics.  This indicates that intensities of bright sources in
uncrowded fields are typically affected by systematic errors of $\sim$3\%.
Such systematic effects are represented in all sources ($>10 \sigma$) by
adding an uncertainty corresponding to 3\% of the intensity in
quadrature with the counting statistical uncertainty.

Other sources may be affected by systematic errors beyond the level
apparent in the Crab Nebula light curve.  First, there are likely to
be additional systematic effects upon the intensity of a source when
other sources, particularly bright ones, are also in the FOV. Our
results for Sco X-1 indicate that other sources in the field may be
affected at the level of a few percent (of Sco X-1). Second,
incompleteness of the active catalog can lead to errors. Third, low
energy absorption by the window and thermal shield increases with
elevation of a source in the FOV. Currently the elevation corrections
are based on the integrated 1.5 - 12 keV Crab flux only. Sources with
markedly different spectra could have additional scatter in the light
curves.  We caution the reader to remain aware of the above
qualifications when using either the results from selected sources,
which we now present, or the comprehensive set of results available
elsewhere.

The light curve of Cygnus X-1 (Fig. 3) shows variability that is more
evident with the time resolution of the measurements from individual
dwells (as shown) than in 1-day averages (not shown). In particular,
the increase to $\sim$1.3 Crab for a brief time ($<$ 1 day) around MJD
50187 is not nearly as apparent in the 1-day averages, where the peak
intensity is $\sim$0.75 Crab. The strengthening to $\sim$2 Crab around
MJD 50220 is apparently a transition to the high state (van der Klis
1995); the spectrum appears to soften as the intensity increases.  A
light curve of much longer duration and a time resolution of days has
been presented by \markcite{pried1983} Priedhorsky, Terrell, \& Holt
(1983).  We note that occasions with the intensity as bright as 1.3
Crab are unusual, at least in time-averages of one day or longer.

The low-mass X-ray binary 4U 1705-44 is seen to vary from $<$25 mCrab to
$\sim$300 mCrab (Fig. 3).  Previous observers have noted variability over a
similar intensity range \markcite{lang1987} (Langmeier et al. 1987, 
\markcite{vanp1995} van Paradijs 1995).
This source exhibits X-ray burst and other aperiodic variability that
appears to be correlated with the intensity state \markcite{lang1987} 
(Langmeier et al. 1987, \markcite{lang1989} Langmeier, Hasinger, \& 
Trumper 1989).

No emission from the X-ray transient GRO J1655-40 above $\sim$12 mCrab was
detected with the ASM prior to MJD 50198, at which time the count rate
rose to $\sim$2 Crab over the course of $\sim$10 days (Fig. 3).  
Outbursts of this source have been
previously detected in 1994 and 1995 \markcite{harm1995} (Harmon et al. 1995, 
\markcite{wilson1995} Wilson et al. 1995, 
\markcite{sazonov1995} Sazonov \& Sunyaev 1995, \markcite{zhang1995} 
Zhang et al. 1995 \& references
therein).  This source is a dynamical black-hole candidate 
\markcite{bailyn1995} (Bailyn et al. 1995) that manifests relativistic radio 
jets \markcite{tingay1995} (Tingay et al. 1995, \markcite{hjell1995} Hjellming 
and Rupen 1995). Optical and radio activity that are correlated with X-ray 
activity have been previously seen \markcite{bailyn1995} (Bailyn et al. 1995, 
\markcite{tingay1995} Tingay et al. 1995).

The SMC X-1 light curve (Fig.  3) indicates that the source was
detected in one-day averages from MJD 50146 to about MJD 50180, and
again starting at about MJD 50202.  This on-off behavior is very
similar to that reported previously by \markcite{gruber1984} Gruber
and Rothschild (1984), who suggested that it may be a $\sim60^{\rm d}$
quasi-periodicity, and if so, would likely be analogous to the $35^{\rm d}$
cycle of Her X-1 and the $30^{\rm d}$ cycle of LMC X-4.

The larger collection of light curves includes many interesting
results that cannot be presented here.  For example, one can see the
gradual decay in the flux of GRO J1744-28, large fluctuations in the
intensities of GRS 1739-278 and GRS 1915+105, the end of an outburst
of 4U 1608-52, and a flux increase in LMC X-3. The long-term
periodicities of Her X-1, LMC X-4, and GX 301-2 \markcite{vanp1995}
(see references in van Paradijs 1995) are also evident.

We have searched each of the $\sim100$ light curves for periodicities.
For this purpose, the measurements from individual dwells were
averaged in 0.05 day bins and the mean count rate was subtracted from
each bin that contained at least one measurement.  The resulting
binned light curves were transformed using a fast Fourier transform to
obtain power density spectra.  The power spectra have been examined
for obviously significant peaks in the period range 0.1 to $\sim$15
days.  The orbital periods for 6 binary systems are clearly apparent
in the results.  These include the orbital periods of Cen X-3
($2.1^{\rm d}$), Cir X-1 ($16.6^{\rm d}$), Cyg X-3 ($4.8^{\rm h}$),
Her X-1 ($1.7^{\rm d}$), Vela X-1 ($9.0^{\rm d}$), and X1700-377
($3.4^{\rm d}$; \markcite{vanp1995} see van Paradijs 1995).  We will
continue to search for periodicities as measurements are added to the
light curves.

Finally, we have estimated typical sensitivities to sources in
uncrowded fields from the scatter in the measurements of 5 sources
whose light curves (not shown) indicate that they have not been
detected. The result is $\sim35$ mCrab ($2 \sigma$) for those individual
dwells in which the source appears in the central half of the field of
view and $\sim10$ mCrab for the one-day averages.

\section{Conclusion}

The ASM is producing interesting and timely results on many X-ray
sources.  We encourage the scientific community and all interested
persons to explore and make use of the ASM results via a public
archive maintained by the {\it RXTE} Guest Observer Facility (see the World
Wide Web URL 
\\http://heasarc.gsfc.nasa.gov/docs/xte/xte\_1st.html
\\for information).

\acknowledgments

We gratefully acknowledge the essential contributions of William Mayer
and Robert Goeke at MIT and also those of every member of the {\it RXTE}
teams at MIT and GSFC to the success of the ASM and {\it RXTE}. The {\it RXTE}
team at MIT was supported by NASA Contract NAS5-30612.

\newpage

\figcaption[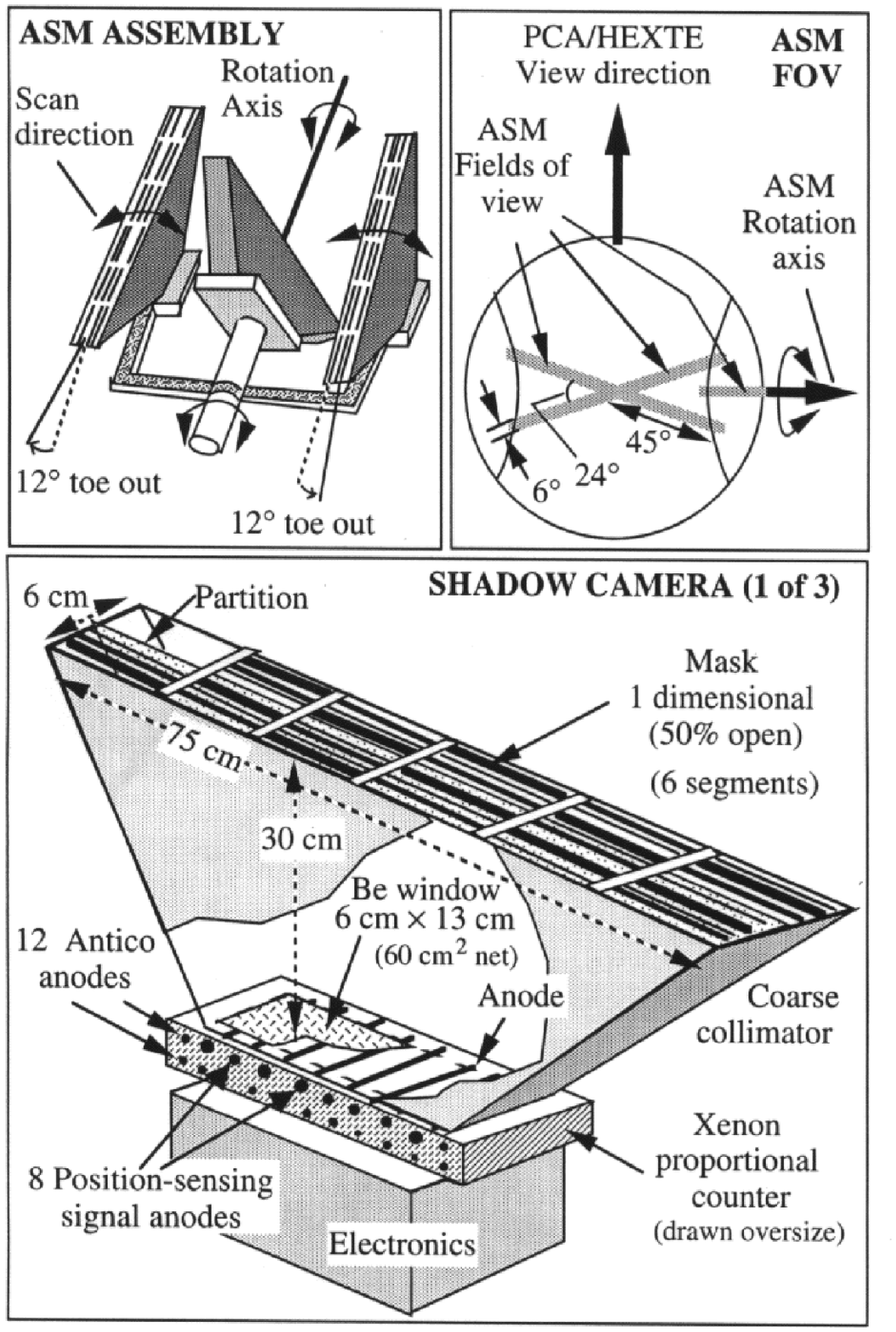]{ {{\it Top left:}} Schematic diagram
showing the relative orientations of the SSCs as configured on the
ASM. {\it Top right:} The centers of the fields of view (FOVs) of SSC
nos. 1 and 2 are approximately coaligned perpendicular to the ASM
rotation axis. The long axes of their FOVs are tilted by +12$\arcdeg$
and -12$\arcdeg$ relative to the rotation axis. The center of the FOV
of SSC no. 3 is pointed parallel to the rotation axis. The view
direction of the other two RXTE instruments is indicated. {\it
Bottom:} Schematic diagram of an SSC illustrating the major
components. One subsection of the mask is actually nearly as long as
the Be window.}

\figcaption[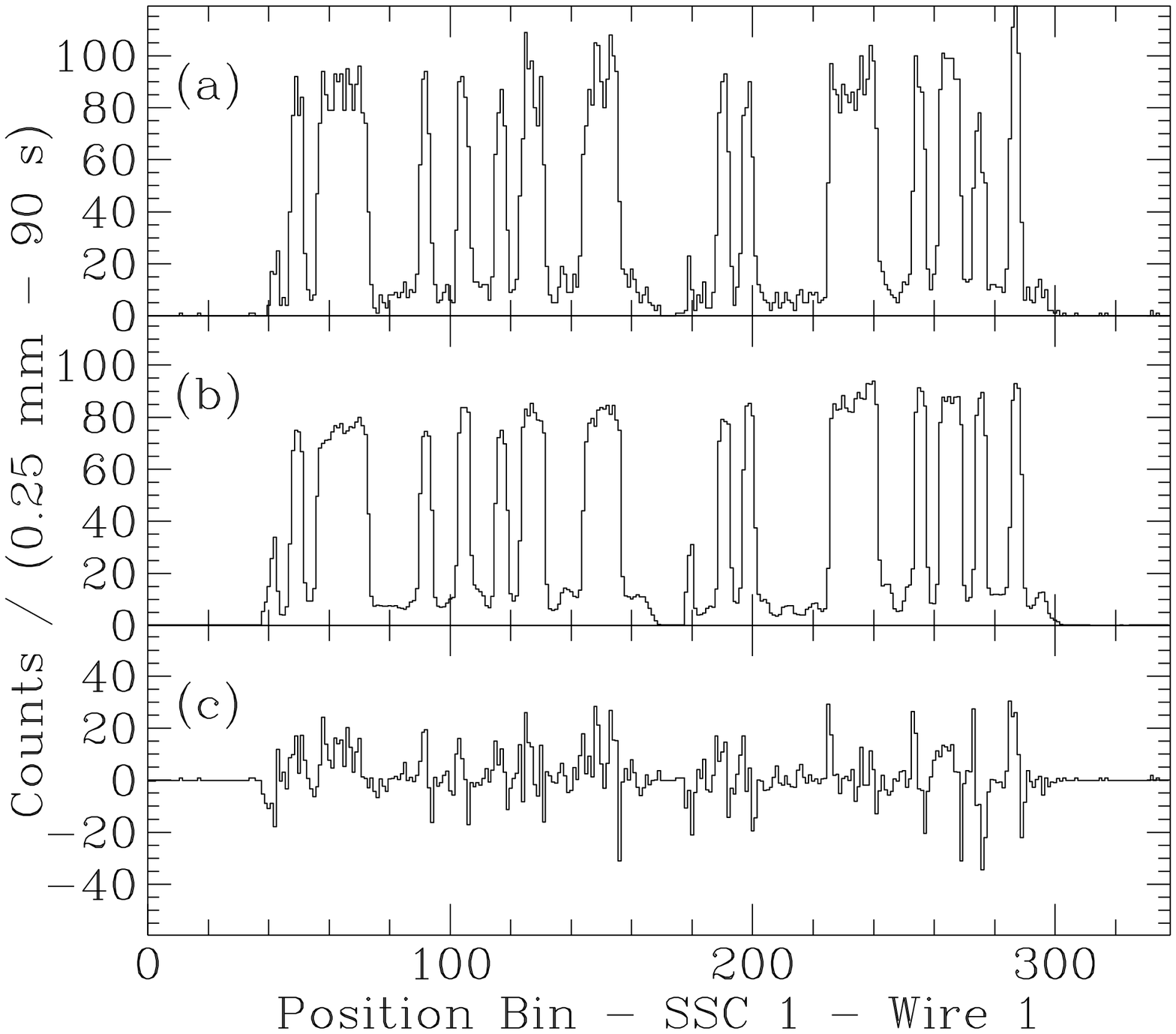]{(a) A position histogram obtained 1996 April 19 
with Sco X-1 at -0.67$\arcdeg$, 23.0$\arcdeg$ from the center of the FOV in 
the narrow and long directions, respectively. There were also 10 other sources 
from the active catalog in the field, but the brightest of these was 
$\sim$7\% as bright as Sco X-1. One position bin corresponds to a region of 
width $\sim$0.25 mm. (b) Model histogram, and (c) Residuals, i.e., the 
difference between the observed data and the model.  The magnitude of the
residuals illustrates a need to improve the histogram model by
refining calibration parameters.  The reduced chi-square statistic for
the fit of the data from all 8 anodes is 1.87.}

\figcaption[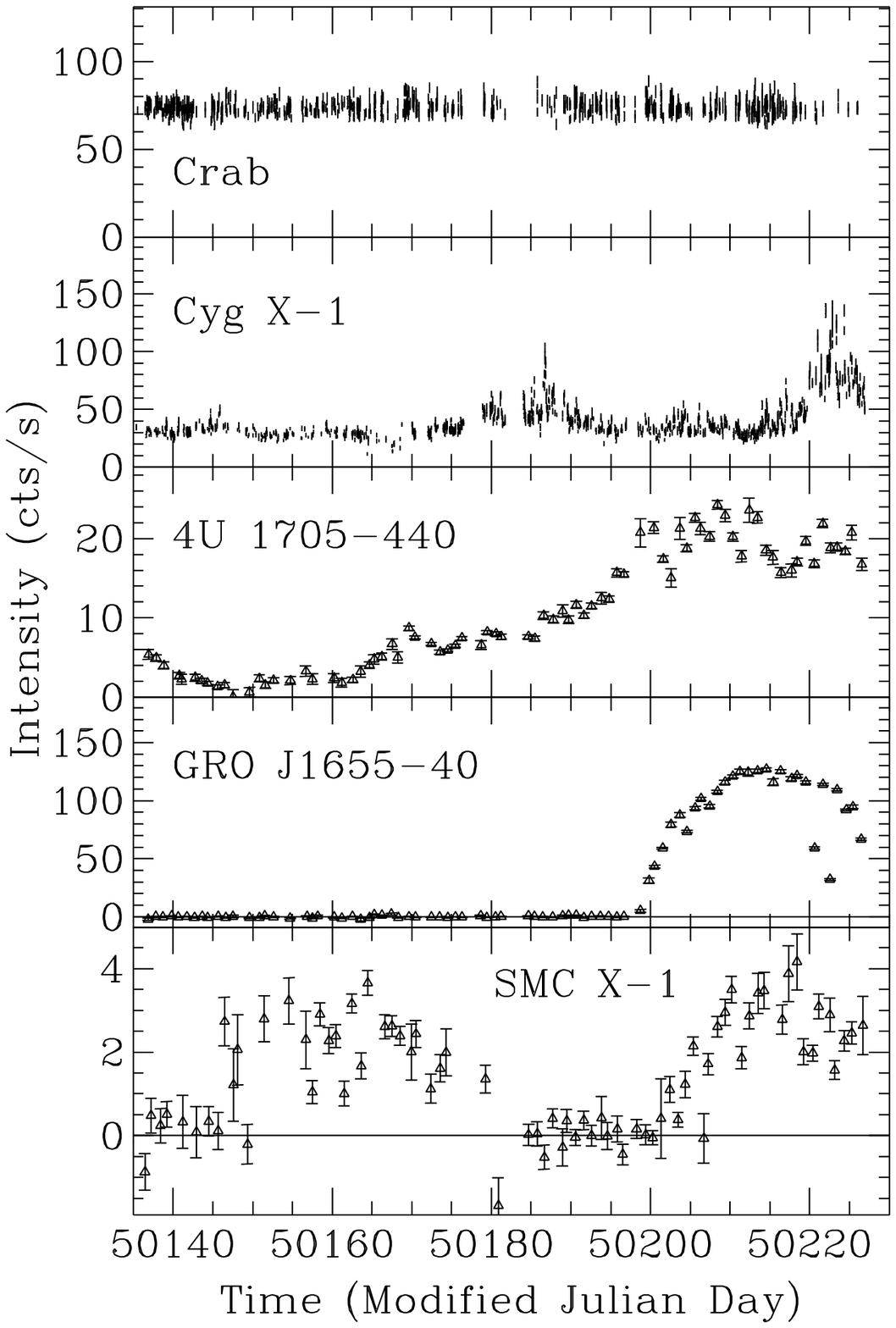]{X-ray light curves (1.5 - 12 keV) for the Crab Nebula, 
Cyg X-1, 4U 1705-44, GRO J1655-40, and SMC X-1. The light curves for the
Crab Nebula and Cyg X-1 comprise measurements from individual dwells.
The bottom three light curves comprise intensities from individual
dwells averaged in one-day time bins.  The error bars represent $\pm1\sigma$
statistical uncertainty added in quadrature with a simple estimate of
systematic errors (see text). MJD 50082.0 is equivalent to 1996 Jan. 0.0 UT.}

\end{document}